\documentclass[10pt]{article}

\usepackage[dvips]{graphicx}
\usepackage{amsmath}
\usepackage{amssymb}

\begin{document}

\begin{center}

\vspace{0.5cm}
\textbf{\Large Uniform WKB approximation of Coulomb wave functions for arbitrary partial wave} 

\vspace{5mm} {\large N.~Michel}

\vspace{3mm}
\textit{CEA, Centre de Saclay, IRFU/Service de Physique Nucl{\'e}aire, F-91191 Gif sur Yvette, France }
\end{center}

\hrule \vspace{5mm} \noindent{\Large \bf Abstract}
\vspace*{5mm}

{Coulomb wave functions are difficult to compute numerically for extremely low energies, even with direct numerical integration.
Hence, it is more convenient to use asymptotic formulas in this region. 
It is the object of this paper to derive analytical asymptotic formulas valid for arbitrary energies and partial waves. 
Moreover, it is possible to extend these formulas for complex values of parameters.
}

\vspace{5mm} \noindent{\bf PACS:}
02.30.Fn, 02.30.Gp, 03.65.Ge

\section{Introduction}
Coulomb wave functions are the typical example of functions being both analytic and very difficult to compute.
They are part of the rare cases for which second-order differential equations can be analytically solved
and are expressed with confluent hypergeometric functions \cite{Abramowitz_Stegun}:
\begin{eqnarray}
F_{\ell \eta} (\rho) &=& C_\ell(\eta) \; \rho^{\ell+1} \; e^{i \omega \rho} \; {_1F_1} \left( 1+\ell+i\omega\eta ; 2\ell+2 ; -2i \omega \rho \right), 
\label{F_analy} \\
H^{\omega}_{\ell \eta} (\rho) &=& e^{i \omega \left[ \rho - \eta \log (2\rho) - \ell \frac{\pi}{2} + \sigma_\ell(\eta) \right]} \; {_2F_0} \left( -\ell+i\omega\eta,1+\ell+i\omega\eta;;-\frac{i}{2 \omega \rho} \right) \label{H_analy}, \\
\sigma_\ell(\eta) &=& \frac{\log \Gamma (1+\ell+i\eta) - \log \Gamma(1+\ell-i\eta)}{2i} \label{sigma_l_eta}, \\
G_{\ell \eta} (\rho) &=& \frac{H^{+}_{\ell \eta} (\rho) + H^{-}_{\ell \eta} (\rho)}{2}  \label{G_from_H}, \\
C_\ell(\eta) &=& 2^\ell \; \exp \left[ -\pi \eta  - \log \Gamma(2\ell+2) \right] \exp \left[ \frac{\log \Gamma(1+\ell+i\eta) + \log \Gamma (1+\ell-i\eta)}{2} \right] \label{Cl_eta} ,
\end{eqnarray}
where $\ell$ is the angular momentum of the wave function, $\eta$ its Sommerfeld parameter, $\omega$ can be equal to $\pm 1$ in {eqs.}(\ref{F_analy},\ref{H_analy}),
and where the constants of normalization $C_\ell(\eta)$ (Gamow factor) and $\sigma_\ell(\eta)$  (Coulomb phase shift) appear \cite{Abramowitz_Stegun}.
All computational difficulty arises from the presence of confluent hypergeometric functions $_1F_1$ and $_2F_0$. They indeed vary by many orders
of magnitude for smooth variations of parameters and are moreover subject to a cut in the complex plane when analytically continued \cite{Abramowitz_Stegun}.
Codes handling arbitrary complex parameters in Coulomb wave function computation 
have been published in both Fortran \cite{Thompson} and more recently in C++ \cite{Coulomb_CPC} languages. While the former uses only analytical methods
such as power series and continued fractions, the latter included direct integration as well, which has considerably extended the numerical domain of definition
for which implementation of Coulomb wave functions is stable \cite{Coulomb_CPC}. However, even with this amelioration, results become unreliable
for very large values of $|$Im$(\ell)|$ and/or $|\eta|$ \cite{Coulomb_CPC}.

In {ref.}\cite{EPJ_Grama}, a uniform approximation for Coulomb wave functions has been presented, which, however, demands $\ell = 0$.
This prevents partial decomposition of wave functions, useful for reaction cross section calculation \cite{Messiah}. 
Moreover, approximations for which $\ell$ is arbitrary would be of interest for relativistic calculations, where Coulomb wave functions are expressed also
with the confluent hypergeometric functions appearing in {eqs.}(\ref{F_analy},\ref{H_analy}) \cite{Seaton}.

\section{Uniform approximation of Coulomb wave functions}
In order to alleviate instabilities encountered in Coulomb wave function implementation,
we will derive analytic formulas valid for large $\eta$, $\ell \geq 0$ and $\rho > 0$ firstly, which will be secondly analytically continued to complex values.
The standard method therein is to use uniform WKB approximation, where divergences occurring at the turning point, denoted in the following as $\rho_t$,
are removed through the use of Airy functions \cite{Berry_Mount}. While the method described in {ref.}\cite{Berry_Mount} demands in the general 
case to deal with non-analytical integrals, those appearing for the Coulomb problem can be calculated exactly with elementary functions.
To apply this method, one firstly writes the exact ansatz verified by Coulomb wave functions, from which approximations can be effected:
\begin{eqnarray}
F_{\ell \eta} (\rho) &=& \sqrt{\pi} \rho_t^{\frac{1}{6}} \phi'(x)^{-\frac{1}{2}} {\mbox{Ai}}(-\rho_t^{\frac{2}{3}} \phi(x)), \label{F_ansatz} \\
G_{\ell \eta} (\rho) &=& \sqrt{\pi} \rho_t^{\frac{1}{6}} \phi'(x)^{-\frac{1}{2}} {\mbox{Bi}}(-\rho_t^{\frac{2}{3}} \phi(x)), \label{G_ansatz} \\
x &=& \frac{\rho - \rho_t}{\rho_t}, \label{x_def} \\
\rho_t &=& \eta + \sqrt{\eta^2 + \ell(\ell+1)}, \label{turning_point}
\end{eqnarray}
where {Ai$(z)$} and {Bi$(z)$} are the standard regular and irregular Airy functions \cite{Abramowitz_Stegun}
(their normalization will be justified afterward, as well as the reason why $F_{\ell \eta}$ ($G_{\ell \eta}$) possesses no {Bi (Ai)} component),
and where $\phi(x)$ verifies the following third-order non-linear differential equation:
\begin{eqnarray}
&&\phi'(x)^2 \phi(x) + \frac{1}{2 \rho_t^2} \phi^{'''}(x) \phi'(x)^{-1} - \frac{3}{4 \rho_t^2} \phi''(x)^2 \phi'(x)^{-2} = \frac{x}{x+1} + \frac{ax}{(x+1)^2}, \label{phi_eq} \\
&&a = 1 - \frac{2 \eta}{\rho_t} \label{a_def},
\end{eqnarray}
where the parameter $a$ has been introduced for convenience. For large $\eta$, $\rho_t \rightarrow +\infty$, so that terms
proportional to $\rho_t^{-2}$ in eq.(\ref{phi_eq}) can be neglected. Hence, $\phi(x)$ verifies asymptotically a non-linear first order equation.
Elimination of turning point divergence moreover demands that $\phi(0) = 0$, so that its approximate equation and solution read for real $x$:
\begin{eqnarray}
\phi'(x)^2 \phi(x) &=& \frac{x}{x+1} + \frac{ax}{(x+1)^2}, \label{phi_app_eq} \\
\frac{2}{3} \phi(x)^{\frac{3}{2}} &=& \int_{0}^{x} \sqrt{\frac{t}{t+1} + \frac{a t}{(t+1)^2}}~dt \mbox{, } x \geq 0, \nonumber \\
\frac{2}{3} (-\phi(x))^{\frac{3}{2}} &=& \int_{0}^{-x} \sqrt{\frac{t}{1-t} + \frac{a t}{(1-t)^2}}~dt \mbox{, } x < 0. 
\label{phi_int_app}
\end{eqnarray}
This approximation has been considered in {ref.}\cite{EPJ_Grama} in the particular case $\ell = 0$, implying $a = 0$, for which $\phi(x)$ is equal to
a particularly simple expression. For $x \rightarrow +\infty$, one can verify easily that $(2/3) \phi(x)^{3/2} \sim x$, which justifies the ansatz
of {eqs.}(\ref{F_ansatz},\ref{G_ansatz}), where they reduce to standard sine-cosine approximation \cite{Abramowitz_Stegun}.
If $a \neq 0$, it happens that eq.(\ref{phi_int_app}) not only can be integrated analytically, but also in a very concise way:
\begin{eqnarray}
\frac{2}{3} \phi(x)^{\frac{3}{2}} &=& (1-a) [\log(\sqrt{1+a}) - \log(\sqrt{x} + \sqrt{1+x+a})] \nonumber \\  
                                  &+& \sqrt{x(1+a+x)} - 2 \sqrt{a} \mbox{ arctan} \left( \sqrt{\frac{ax}{1+a+x}} \right)  
                                  \mbox{, }  x \geq 0, \nonumber \\
\frac{2}{3} (-\phi(x))^{\frac{3}{2}} &=& -\sqrt{-x(1+a+x)} 
                                 + \frac{1-a}{2} \mbox{arccos} \left( 1 + \frac{2x}{1+a} \right) \nonumber \\ 
                                 &+& 2 \sqrt{a} \mbox{ arctanh} \left( \sqrt{-\frac{ax}{1+a+x}}  \right)
                                 \mbox{, } x < 0. \label{phi_analy_app}
\end{eqnarray}
One can check that the $\ell = 0$ case described in {ref.}\cite{EPJ_Grama}
is properly obtained using $a = 0$ in eq.(\ref{phi_analy_app}).
$\phi'(x)$ is obtained by eq.(\ref{phi_app_eq}) and the condition that $\phi'(x) > 0$,
immediate from eq.(\ref{phi_int_app}). Note that for $x \sim 0$, 
eq.(\ref{phi_analy_app}) becomes numerically unstable, so that it is preferrable
to use the asymptotic formulas verified by $\phi(x)$ and $\phi'(x)$ therein, which are $\phi(x) \sim (1+a)^{1/3} x$ and $\phi'(x) \sim (1+a)^{1/3}$.
$F'_{\ell \eta} (\rho)$ and $G'_{\ell \eta} (\rho)$ are obtained by a simple differentiation of {eqs.}(\ref{F_ansatz},\ref{G_ansatz}).
It has been noticed numerically, however, that their term proportional to $\phi''(x)$ is not negligible and should be kept in approximate formula.
$\phi''(x)$ is obtained by a simple differentiation of eq.(\ref{phi_app_eq}).

Extension of eq.(\ref{phi_analy_app}) to complex arguments demands caution.
On the one hand, it is usually sufficient therein to replace the conditions $x \geq 0$ and $x < 0$
by Re$(x) \geq 0$ and Re$(x) < 0$ respectively, especially if considered imaginary parts are small in modulus 
(see also {ref.}\cite{Airy} for computational methods of the Airy function in the complex plane).
On the other hand, however, one has to pay attention to the different cuts obeyed by both Coulomb wave functions
and the elementary functions of eq.(\ref{phi_analy_app}). 
{The theoretical behavior of Coulomb wave functions in the vicinity of their cut 
has been studied in {ref.}\cite{Dzieciol}.}
The simplest method to avoid problems generated by cuts 
is to consider complex contours which never cross the negative real axis, so that Coulomb wave functions are
continuous therein. Then, one just has to modify formulas of eq.(\ref{phi_analy_app}) 
so that they are continuous on these contours if cuts of elementary functions therein are encountered.
They are straightforward to treat, as cuts appear only by way of $\log$, inverse {circular/hyperbolic} and power functions.

\section{Numerical examples}
\begin{figure}[htb]
\includegraphics[angle=-90,width=1\textwidth]{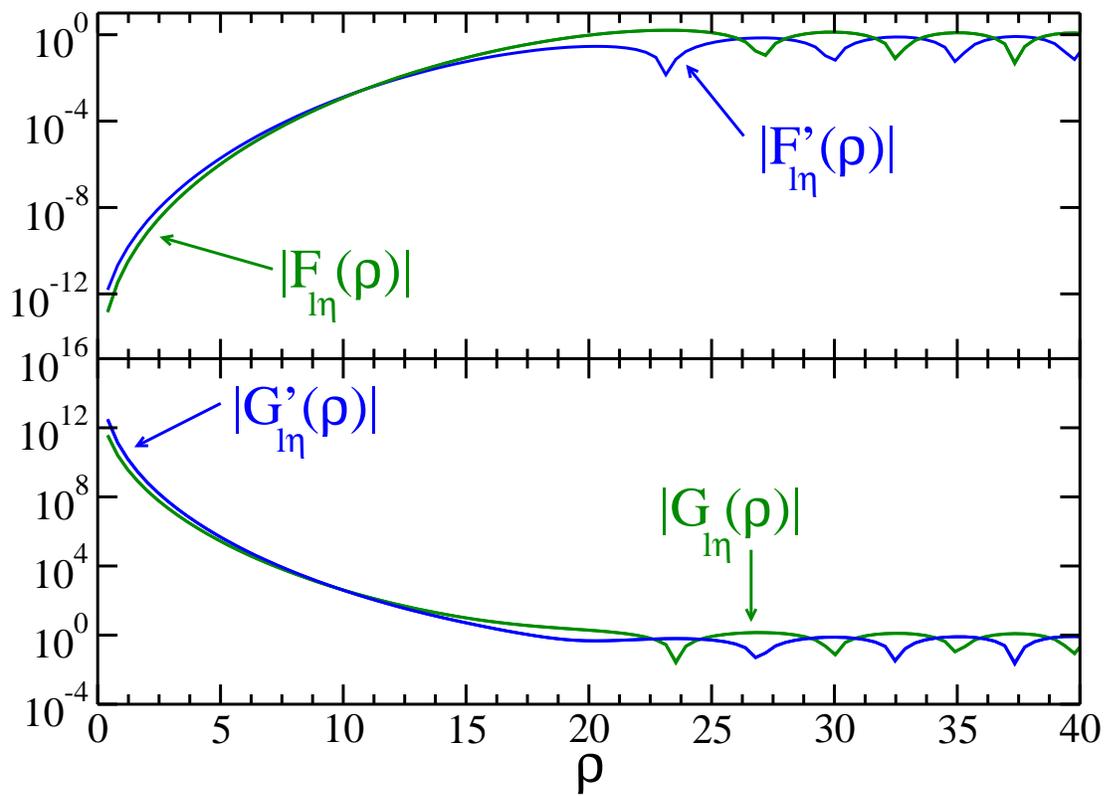}
\protect\caption{\label{real_fig} 
Absolute values of $F_{\ell \eta} (\rho)$, $G_{\ell \eta} (\rho)$, $F'_{\ell \eta} (\rho)$ and $G'_{\ell \eta} (\rho)$ for $\ell = 2$ and $\eta = 10$.
Exact calculation provided by the code of {ref.}\cite{Coulomb_CPC} and approximation issued from eq.(\ref{phi_analy_app}) are indistinguishable.
}
\end{figure}

\begin{figure}[htb]
\includegraphics[angle=-90,width=1\textwidth]{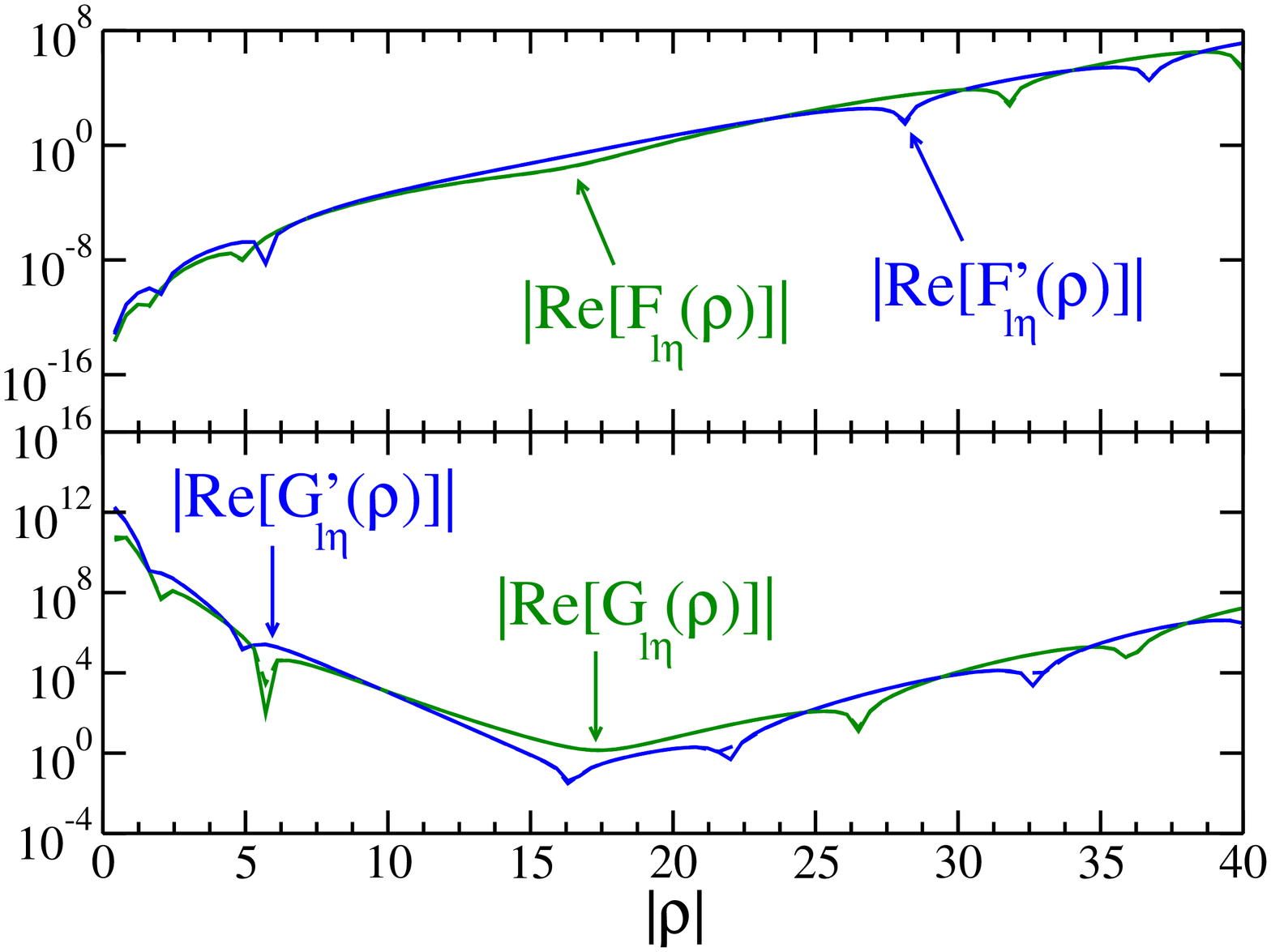}
\protect\caption{\label{complex_fig}
Absolute values of Re$[F_{\ell \eta} (\rho)]$, {Re$[G_{\ell \eta} (\rho)]$,} Re$[F'_{\ell \eta} (\rho)]$ and Re$[G'_{\ell \eta} (\rho)]$ 
for $\ell = 2 + i$, $\eta = 10 + i$ and arg$(\rho) = \pi/4$.
Exact calculation provided by the code of {ref.}\cite{Coulomb_CPC} is provided as straight lines
and approximation issued from eq.(\ref{phi_analy_app}) as dashed lines. 
Differences are visible only for $|$Re$[G_{\ell \eta} (\rho)]|$  and $|$Re$[G'_{\ell \eta} (\rho)]|$.
}
\end{figure}

In order to show the efficiency of the approximation presented in eq.(\ref{phi_analy_app}), one will consider both sets of parameters.
The first set consists in the real values $\ell = 2$ and $\eta = 10$, for which $\rho > 0$. The second set reads $\ell = 2 + i$ and $\eta = 10 + i$,
with $\rho = |\rho| e^{i\pi/4}$. No cut problem appears for these complex values in eq.(\ref{phi_analy_app}). 
$|$Re$[F_{\ell \eta} (\rho)]|$, $|$Re$[G_{\ell \eta} (\rho)]|$
and analog values related to Coulomb wave functions derivatives are depicted in log scale, in fig.(\ref{real_fig})
for the real set of parameters and in fig.(\ref{complex_fig}) for the complex set of parameters.
They are compared to the exact functions calculated numerically with the code of {ref.}\cite{Coulomb_CPC}, with which Airy functions are calculated as well
as linear combinations of Bessel functions \cite{Abramowitz_Stegun}.
It is clear from these figures that the approximation provided by eq.(\ref{phi_analy_app}) is very good, even though $\eta$ is not very large.
The relative error of the approximate formula of eq.(\ref{phi_analy_app}) is $\sim$1\% for most $\rho$ values, which was expected
as precision of the approximation is of the order of $|\rho_t|^2$ (see eq.(\ref{phi_eq})).

\section{Conclusion}
The uniform Coulomb wave function approximation of {ref.}\cite{EPJ_Grama}, valid for $\eta \rightarrow +\infty$ and $\ell = 0$,
has been generalized to both arbitrary partial wave and complex parameters, and has been checked numerically to be reliable.
It provides a useful alternative to exact computation of low-energy Coulomb wave functions, which can be numerically costly
or unstable. In particular, partial wave decomposition can be effected with the proposed 
uniform approximation of low-energy wave functions.

\end{document}